\documentclass[runningheads]{svmult}

\usepackage{makeidx}   
\usepackage{graphicx}  
\usepackage{subeqnar}  
\usepackage{multicol}  
\usepackage{physprbb}  
\makeindex             

\catcode`\@=11
\def\gsim{\ifmmode{\mathrel{\mathpalette\@versim>}}
    \else{$\mathrel{\mathpalette\@versim>}$}\fi}
\def\lsim{\ifmmode{\mathrel{\mathpalette\@versim<}}
    \else{$\mathrel{\mathpalette\@versim<}$}\fi}
\def\@versim#1#2{\lower 2.9truept \vbox{\baselineskip 0pt \lineskip 
    0.5truept \ialign{$\m@th#1\hfil##\hfil$\crcr#2\crcr\sim\crcr}}}
\catcode`\@=12
\def\sg0{\sigma_0}
\def\mbh{M_{\rm BH}}
\def\Lb{L_{\rm B}}

\def\cRe{\langle R\rangle _{\rm e}}
\begin{document}
\title*{Galaxy merging and the Fundamental
\protect\newline Plane of elliptical galaxies}
\toctitle{Galaxy merging and the Fundamental Plane
\protect\newline of elliptical galaxies}
%
%
\titlerunning{Galaxy merging and the FP}
%
\author{Carlo Nipoti\inst{1}
\and Pasquale Londrillo\inst{2}
\and Luca Ciotti\inst{2,3}}
\authorrunning{Carlo Nipoti et al.}
%
%
\institute{Dipartimento di Astronomia dell'Universit\`a degli Studi di 
           Bologna, via Ranzani 1, 40127 Bologna, Italy
\and Osservatorio Astronomico di Bologna, via Ranzani 1, 40127 Bologna, Italy 
\and Scuola Normale Superiore, Piazza dei Cavalieri 7, 56126 Pisa, Italy }

\maketitle              

\begin{abstract}
We present preliminary results of numerical simulations of
dissipationless merging of stellar systems, aimed at exploring the
consequences of merging between gas free, spheroidal systems. In
particular, we study the dynamical and structural characteristics of
hierarchical merging between equal mass stellar systems, and we
compare the properties of the end-products with the most important
structural and dynamical scaling relations obeyed by spheroids. In the
explored hierarchy of four successive mergings we find that the FP
tilt is marginally conserved, but both the Faber-Jackson and Kormendy
relations are {\it not} conserved.
\end{abstract}

\section{Introduction}
From a {\it theoretical} point of view, in the scenario of
hierarchical galaxy formation elliptical galaxies (Es) formed by
merging of smaller systems~\cite{4,11,12}. On the other hand, from
{\it observations} we know that Es satisfy many tight scaling
relations: for example the Fundamental Plane~\cite{5,6} (FP), the
$\mbh-\sg0$~\cite{8,9}, the Mg$_2-\sg0$~\cite{1}, and the
color--magnitude~\cite{2} relations. In particular, the FP of Es
relates their central velocity dispersion $\sg0^2$, total luminosity
$\Lb$ and circularized effective radius $\cRe$, with a 1-sigma scatter
of $\simeq15\%$ in $\cRe$ for fixed $\Lb$ and $\sg0$~\cite{10}.

{\it Here we focus on the constraints imposed by the existence of the
FP on the role of dissipationless merging in the formation of Es. In
other words we want to verify, by using numerical simulations, whether
the FP is ``closed'' with respect to the merging process}.

Among the motivations of this exploration is the fact that, in the
merging of two galaxies with masses $(M_1 ,M_2)$ and virial velocity
dispersions $(\sigma_{v,1},\sigma_{v,2})$, the virial velocity
dispersion of the merger (in case of no mass loss and negligible
initial interaction energy of the galaxy pair when compared to their
internal energies) is given by
\begin{equation}
{\sigma^2_{v,1+2}}={M_1{\sigma^2_{v,1}}+M_2{\sigma^2_{v,2}}\over{M_1+M_2}}.
\end{equation} 
It follows that $\sigma_{v,1+2}\leq\,{\rm
max}(\sigma_{v,1},\sigma_{v,2})$, i.e., {\it the virial velocity
dispersion cannot increase in a merging process of the kind described
above}.  On the other hand, the Faber-Jackson relation~\cite{7} (FJ)
indicates that the {\it projected central velocity dispersion}
increases with galaxy luminosity.  In addition, since in a purely gas
free merging process the stellar mass--to--light ratio cannot
increase, {\it the FP can be maintained only by structural and/or
dynamical non--homology}.

\section{Results}
The details of the adopted (one and two component) numerical models
are given in~\cite{14}.  We describe here the case of a {\it merging
hierarchy}: in other words, we merge together the end--products of
previous mergers up to four generations, for a total increase of mass
of a factor of 16.
\begin{itemize}
\item As expected, in all the merging events $\sigma_{v,1+2}$ does not
differ significantly from those of the progenitors (the largest
deviation, due to particle escape, is less than $4\%$).

\item On the contrary, the luminosity-weighted projected velocity
dispersion inside $\cRe/8$ of the end--products $\sigma_a(\cRe/8)$ (an
estimate of the observed $\sg0$), is larger than in the progenitors,
while $\sigma_a(\infty)\simeq \sigma_{v,1+2}$, in accordance with the
projected virial theorem\cite{3}.

\item In general the end--products have $\sg0$ {\it lower} and $\cRe$
{\it larger} than those predicted by the FJ and Kormendy~\cite{13}
relations, respectively. The effects curiously compensate and the
end--products remain near the FP: however, the scatter in $\cRe$ can
be as large as $35\%$, when compared to $\cRe$ derived from the FP
relation when using the total luminosity (mass) and central velocity
dispersion of the end--products.
\end{itemize}
We are now running high-resolution numerical simulations analogous to
those described here, in order to exclude numerical effects on the
above results, and to quantitatively check the effects of massive dark
matter halos on the properties of the end-products.

\end{document}